\documentclass[aps,pre,twocolumn,superscriptaddress,showpacs,showkeys,notitlepage]{revtex4-1}
\usepackage{color}
\usepackage{amssymb}
\usepackage{amsmath}
\usepackage{bm}
\usepackage{graphicx}
\usepackage{epstopdf}
\usepackage{newtxtext,newtxmath}
\usepackage{xfrac}
\usepackage{amsfonts}

\usepackage{bbold}

\newcommand{\ignore}[1]{}
\newcommand{\gcr}{g_{{\rm cmp}\to {\rm rxn}}}
\newcommand{\grc}{g_{{\rm rxn}\to {\rm cmp}}}

\begin{document}
\title{Correlation-enhanced viable core in metabolic networks}
\author{Mi Jin  \surname{Lee}}
\affiliation{Department of Applied Physics, Hanyang University, Ansan 15588, Korea}
\author{Sudo \surname{Yi}}
\affiliation{School of Computational Sciences, Korea Institute for Advanced Study, Seoul 02455, Korea}
\author{Deok-Sun \surname{Lee}}
\email{deoksunlee@kias.re.kr}
\affiliation{School of Computational Sciences, Korea Institute for Advanced Study, Seoul 02455, Korea}
\affiliation{Center for AI and Natural Sciences, Korea Institute for Advanced Study, Seoul 02455, Korea}

\begin{abstract}
Cellular ingredient concentrations can be stabilized by adjusting generation and consumption rates through multiple pathways. To explore the portion of cellular metabolism equipped with multiple pathways, we categorize individual metabolic reactions and compounds as viable or inviable: A compound is viable if processed by two or more reactions, and a reaction is viable if all of its substrates and products are viable. Using this classification, we identify the maximal subnetwork of viable nodes,  referred to as the {\it viable core}, in bipartite metabolic networks across thousands of species. The obtained viable cores are remarkably larger than those in degree-preserving randomized networks, while their broad degree distributions commonly enable the viable cores to shrink gradually as reaction nodes are deleted.  We demonstrate that the positive degree-degree correlations of the empirical networks may underlie the enlarged viable cores compared to the randomized networks. By investigating the relation between degree and cross-species frequency of metabolic compounds and reactions, we elucidate the evolutionary origin of the correlations. 
\end{abstract}

\date{\today}

\maketitle
\section{Introduction}

Living organisms adapt to fluctuating environments over generations, resulting in the evolution of cellular metabolism through the creation, deletion, and modification of metabolic pathways~\cite{Horowitz153,Ycas:1974dp,doi:10.1146/annurev.mi.30.100176.002205,RN74}.  The stochastic nature of evolution, along with different environments that ancestor species have been exposed to, creates differences among contemporary species. On the other hand, metabolism should be able to execute common key functions essential to life via a common subset of metabolic pathways. Consequently, metabolic reactions (enzymes) and compounds exhibit heterogeneity in the number of species containing them in the metabolism~\cite{Bernhardsson:2011ai,Kim:2015aa,Kim:2019aa} and functional robustness~\cite{10.1093/bioinformatics/btg386,10.1371/journal.pcbi.0010068,Grimbs:2007aa,doi:10.1073/pnas.0803571105}; Some are in charge of key functions and thus well preserved across species and stable in activity against environmental perturbations while others are variable, interacting closely with the environment~\cite{10.1371/journal.pcbi.1000613, RN57}. Living organisms should allocate finite resources to diverse cellular functions optimally towards the maximal growth and reproduction~\cite{Scott1099},  and thus identifying stable and variable part of metabolism can reveal the principle of metabolic resource allocation under the tension between robustness and adaptability~\cite{doi:10.1073/pnas.0803571105}.

A stable component of metabolism should consist of the compounds that can maintain steady concentrations against possible perturbations by balancing generation and consumption rates~\cite{https://doi.org/10.1002/bit.260260210,10.1042/bj2380781}. To identify the stable component, one may need to examine the fluxes of all metabolic reactions in all possible conditions, which is however hard experimentally though numerical simulations could be available in limited cases~\cite{doi:10.1128/jb.01743-08}. Here we represent metabolism of each species as a bipartite network connecting reaction nodes to their substrate and product compound nodes by undirected links, assuming that all reactions are reversible,  and obtain the stable component by applying a structural criterion. Such a network representation of metabolism~\cite{Jeong:2000wc}  has been widely used, successfully illustrating key principles of cellular metabolism organization~\cite{pmic.200400962,FRANCKE2005550,RN70,GIANCHANDANI2006284,doi:10.1073/pnas.0802208105,1742-5468-2010-12-P12015}.
In this work we classify individual reaction and compound nodes into viable and inviable ones based on their local connectivity pattern relevant to supporting dynamic stability,  applying the criteria introduced in Refs.~\cite{10.1093/bioinformatics/btg386,doi:10.1073/pnas.0803571105} and suited for our undirected bipartite metabolic networks. Then we define the maximal induced-subgraph consisting of these viable nodes as the {\it viable core}, which can be a candidate for the stable component. The procedures to prune inviable nodes and obtain the viable core are the bipartite-network counterpart  of the greedy leaf removal process considered for a unipartite network in Ref.~\cite{PhysRevLett.109.205703}.

 Our study shows that more (less) resources are allocated for robustness (adaptability) in the real metabolic networks than expected based only on the node degrees. The viable cores of 5469  species from the BioCyc database~\cite{10.1093/bib/bbx085} are significantly larger than those in the randomized networks sharing the same degree sequences.  Individual reactions show different likelihood to belong to the core; Some reactions belong to the core in all species while others do so in just few species. The functional form of the degree distributions can drastically alter  how the core shrinks as reactions are removed randomly, which is revealed by the analytic solutions to the viable cores of uncorrelated networks.  By generating artificial correlated networks, we find that a positive correlation between the degrees of connected nodes 
 can enlarge the viable core, and to understand it, we investigate how the degree-degree correlation affects the branching ratio of inviable nodes in the pruning process. Indeed, positive degree-degree correlations exist in the real metabolic networks, supporting their large cores. Investigating the relation between the cross-species frequencies and degrees of metabolic reactions, we discuss how living species can acquire the degree-degree correlations during evolution. 

This paper is organized as follows. In Sec.~\ref{sec:cores}, we define the viable core of a metabolic network and the viability of a node, and study them in the empirical and randomized networks. The effects of correlations on the core size are investigated, along with the cross-species properties that offer an evolutionary perspective in Sec.~\ref{sec:correlation}. We summarize and discuss the implications of our results in Sec.~\ref{sec:summary}.

\section{Viable core and viability}
\label{sec:cores}
We construct the bipartite metabolic networks for the species included in the version 19.1 of the BioCyc database~\cite{10.1093/bib/bbx085}, covering a total of 5469 bacterial species, 7695 distinct compounds, and 12077 reactions. A species' network is made by connecting each chemical reaction present in the species to its associated compounds, substrates or products; for instance, if a reaction $r$ converts two substrates $c_1$ and $c_2$ into two products $c_3$ and $c_4$, i.e., $c_1+c_2\to c_3+c_4$, then we assign four undirected links $(r,c_1), (r,c_2), (r,c_3)$, and $(r,c_4)$~\cite{Jeong:2000wc,metabolic_network}.  Considering the possibility of the reactions' direction to be reversed depending on the environment and the concentrations of substrates and products, we consider all reactions as reversible and thus assign undirected links. We find $1393 \pm 454$ reaction nodes and $1499 \pm 437$ compound nodes in a network. The number of neighbor nodes, i.e., the connected reactions (compounds) of a compound (reaction), is called degree and denoted by $k$ ($q$).  

\begin{figure}
\includegraphics[width=\columnwidth]{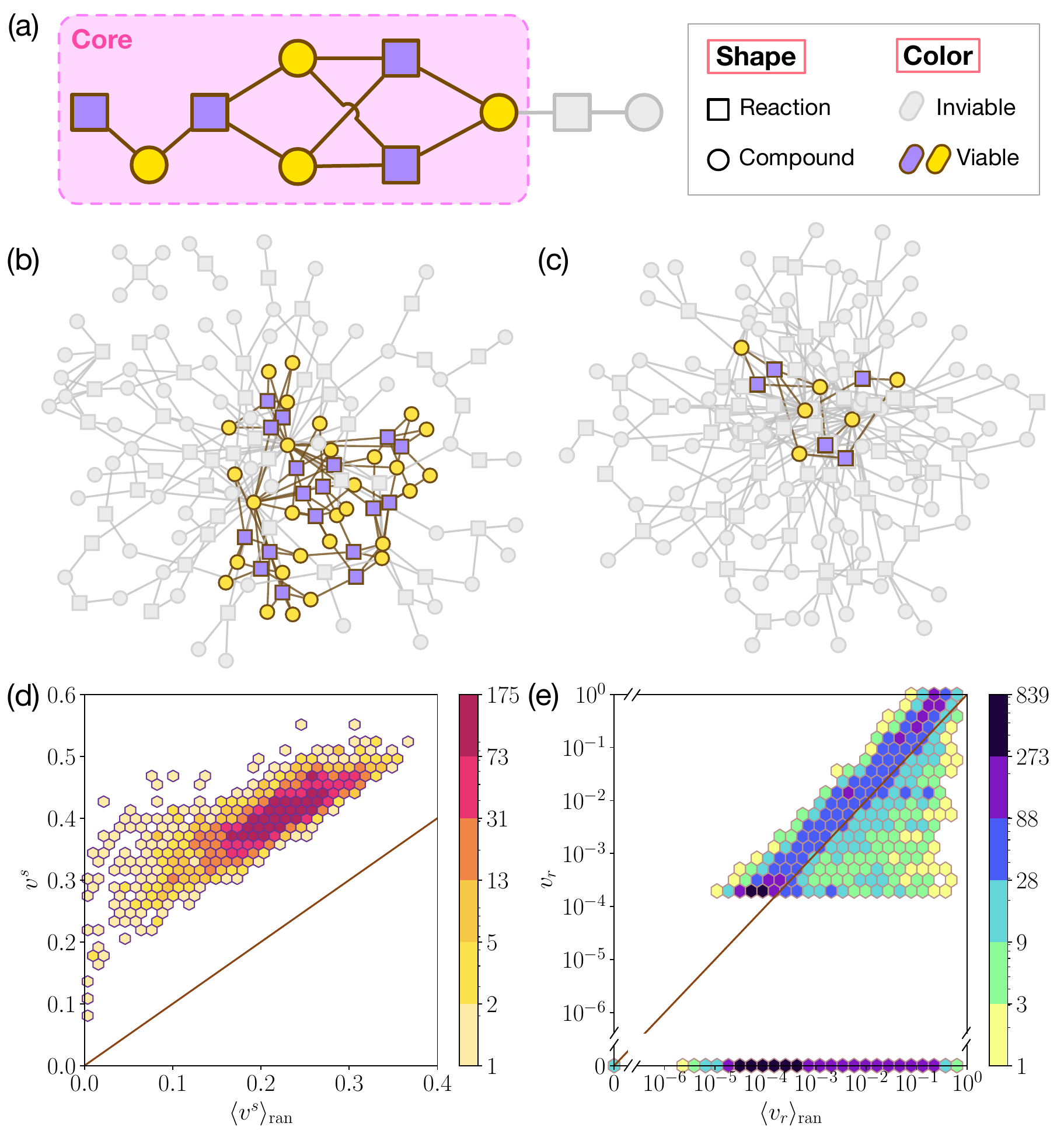}
\caption{Viable cores of metabolic networks. 
(a) An example of the viable core consisting of viable nodes in a bipartite metabolic network. The two rightmost nodes are inviable, which are pruned to identify the core.
The viable core (b) of {\it Klebsiella pneumoniae} and  (c) of its degree-preserving randomized network. 
(d) The relative size of the viable core $v^s$ for 5469 species and (e)  the reaction viability $v_r$ for 12077 reactions. They are plotted versus the counterparts averaged over  100 instances of degree-preserving randomized networks for each species. Color indicates the number of data points. The solid lines are $y=x$ for a guide.}
\label{fig:cores}
\end{figure}

\subsection{Definition and computation of viable core}

Given our goal of characterizing a stable subset of metabolism by a structural criterion, we exploit the fact that a compound can maintain constant concentration only if at least one reaction produces it and another consumes it, which can support adjusting generation and consumption rates respectively, as introduced in Refs.~\cite{10.1093/bioinformatics/btg386,doi:10.1073/pnas.0803571105}. Translating this condition for our undirected networks, we define a compound $c$ as viable in a given network if its degree $k_c$, the number of reactions involving it and thus connected to it, is equal to or larger than two ($k_c\geq 2$); one reaction can generate and another can consume the compound under the assumption that the reactions are reversible. If a compound has no or just a single reaction as its neighbor, then it is considered as inviable. A reaction node $r$ is defined as viable if all of its $q_r$ connected compound nodes are viable; A reaction can occur only if none of its substrates and products are inviable. We present the examples of viable and inviable nodes in  Fig.~\ref{fig:cores}(a). 

Then one can obtain the maximal induced-subgraph consisting of such viable compounds and viable reactions, which we define as the viable core. One can consider the viable core as the subset of metabolism which can maintain constant fluxes and concentrations of the reactions and compounds involved once they are established; Whether the viable core will reach such a steady state depends on the initial condition and the environment. 

The viable core of a network can be obtained by removing iteratively all inviable nodes as follows. We first remove inviable compounds, i.e., the compound nodes of degree smaller than two. Their connected reaction nodes then become inviable and are removed. After this removal of inviable compounds and reactions in the first step, there can appear new inviable compounds, which had degree $k\geq 2$ but now have $k<2$ due to the loss of some of their connected reactions in the previous step. These new inviable compound nodes and their connected reaction nodes are removed in the second step.  Such pruning is repeated until there are no more inviable nodes left, when the remaining compound and reaction nodes constitute the viable core.   An example is presented in Fig.~\ref{fig:cores}(a). This pruning process is almost the same as the greedy leaf removal (GLR) process considered in Ref.~\cite{PhysRevLett.109.205703}, in which the nodes of degree smaller than two and all of their neighbor nodes are removed iteratively in unipartite networks without distinguishing the type of nodes.

\subsection{Enhanced viable core and viability}

Applying the procedure described in the previous subsection, we have obtained the viable core of the metabolic network for each species. As an example, the viable core of \textit{Klebsiella pneumoniae} is shown in Fig.~\ref{fig:cores}(b), which includes 18 viable reactions and 29 viable compounds among a total of 62 reactions and 91 compounds. To understand how large or small the viable core of a species $s$ is, we consider its relative size $v^{s}$, the ratio of the number of nodes in the viable core to the total number of nodes. We compare it with the average relative size $\langle v^{s}\rangle_{\rm ran}$ of the viable cores of the randomized networks,  obtained by rewiring links while preserving the degrees of individual nodes; Each instance of these randomized networks, shown in Fig.~\ref{fig:cores}(c) as an example, has been created by swapping $L^s$ times the end nodes of two randomly chosen links ($L^s$ is the total number of links for species $s$) and we have generated one hundred such instances for each species. One can see that the viable core of the real network of \textit{Klebsiella pneumoniae} [Fig.~\ref{fig:cores}(b)] is much larger than that of a randomized network [Fig.~\ref{fig:cores}(c)]. This  holds for all species; The empirical core size 
ranges between 0.1 and 0.5 across species, and  $v^{s}> \langle v^{s}\rangle_{\rm ran}$ (P-value less than $6\times 10^{-6}$) for every species $s$  as shown in Fig.~\ref{fig:cores}(d). This result means that the core is larger than expected based on the given degree sequences when others are random, and suggests that the network structure beyond the degree sequence plays a role in forming the viable core. Such enhanced stability has been reported in different contexts for a few selected species, e.g., it has been shown that the cascades of failures are more constrained in the real metabolic networks than in the randomized networks~\cite{doi:10.1073/pnas.0803571105,Guell201439}. Also, as we will show below, highly heterogeneous degrees of compounds~\cite{Jeong:2000wc} strongly influence the size of the viable core. 

We define the viability of each reaction $v_r$ as the fraction of the species having the reaction in their viable cores. It can quantify the likelihood that a reaction belongs to the viable core and is thus  stable.  In Fig.~\ref{fig:cores}(e), it is shown that a significant number of reactions ($5549/12077\simeq 46\%$)  do not belong to the viable core in any species, while their viability in the randomized networks is mostly non-zero and distributed broadly---5530 reactions have $v_r = 0$ but $\langle v_r\rangle_{\rm ran}>0$. 
On the other hand,  the other 6528 reactions have non-zero viability in the empirical networks and tend to have lower viability in the degree-preserving randomized networks, i.e., $v_r>\langle v_r\rangle_{\rm ran}$. These findings indicate the stronger heterogeneity of reaction's viability in empirical networks than in randomized networks and imply that the viable cores of real metabolic networks are primarily formed by selected high-viability reactions.

\subsection{Viable core under random failure}

\begin{figure}
\includegraphics[width=\columnwidth]{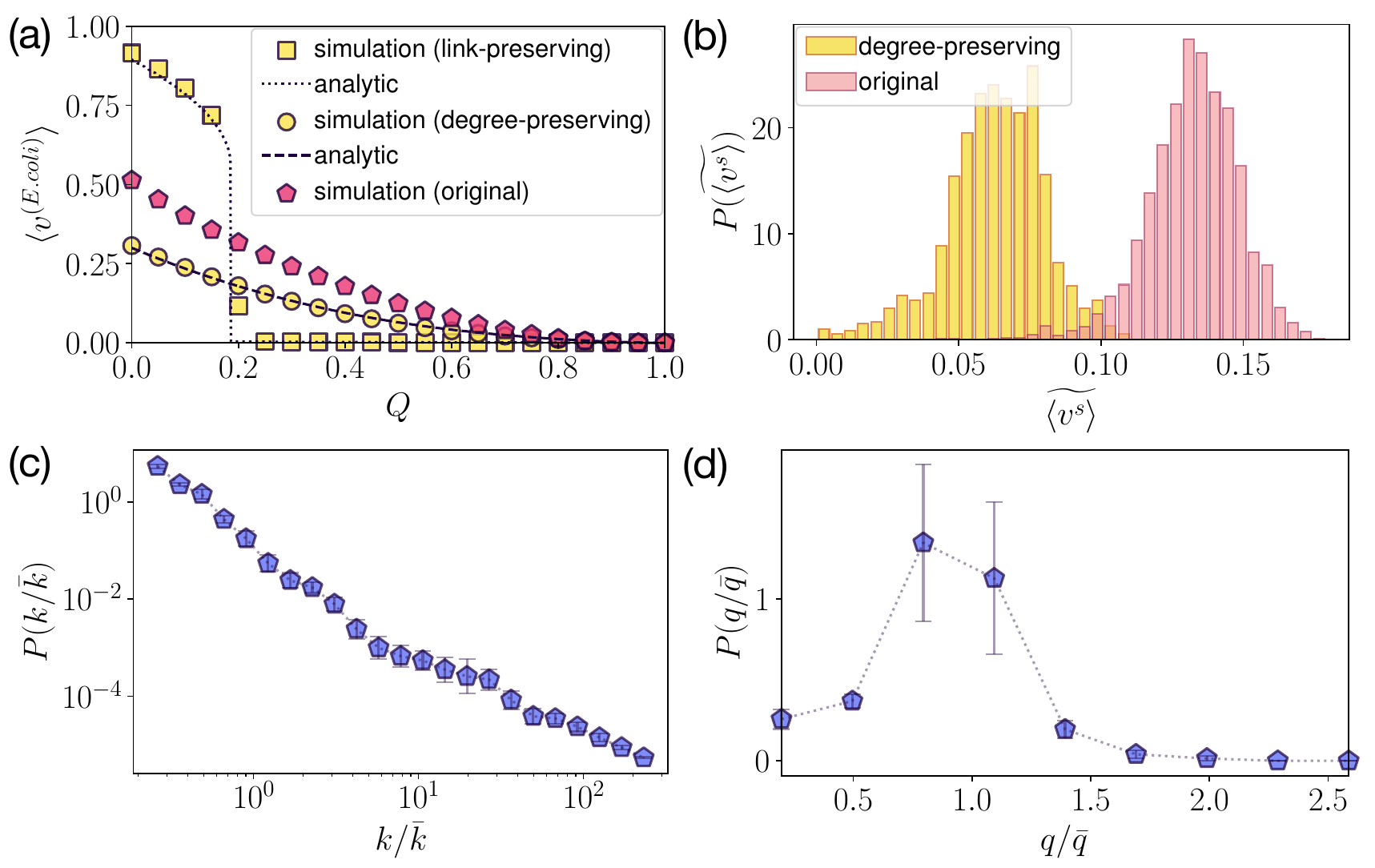}
\caption{
Viable core under random failure. 
(a) Size of the viable core as a function of the fraction $Q$ of removed reaction nodes in the original, degree-preserving randomized, and link-preserving randomized networks for \textit{Escherichia coli}. Lines represent the analytic results from Eq.~\eqref{eq:c}. Error bars are smaller than the data points. 
(b) Distribution of the characteristic size of the viable core under failure for the original networks of all species is located in the right of that for the degree-preserving randomized networks. Distribution of the degree of (c) compounds and of (d) reactions are shown, scaled by the mean degrees $\bar{k}^s$ and $\bar{q}^s$ and  averaged over all species $s$.
}
\label{fig:robust}
\end{figure}

We have shown that the real-world metabolic networks have larger stable components than expected for the randomized networks. 
Will it hold also even with perturbations? In terms of the viable core, we can seek the answer by measuring the viable core in the damaged networks. 

To be specific, we remove a fraction $Q$ of randomly chosen reaction nodes to mimic their malfunction due to e.g., the inactivation or mutation of the enzymes catalyzing them, and then compute the viable core. For \textit{Escherichia coli} [Fig.~\ref{fig:robust}(a)], the real damaged networks have larger cores than  the damaged degree-preserving randomized networks in the whole range of $Q$. For all species $s$, we measure $v^s(Q)$  and $\langle v^s\rangle_{\rm ran}(Q)$ as functions of $Q$ in the original and degree-preserving randomized networks, respectively. We  find that the characteristic core size $\widetilde{v^s} \equiv \int_0^1 v^s (Q) dQ$, an average of $v$ in the range $0\leq Q\leq 1$, of the real damaged networks is much larger than that  of the damaged randomized networks $\widetilde{\langle v^s\rangle}_{\rm ran}$ as shown by comparing their distributions over species [Fig.~\ref{fig:robust}(b)]. 

Since the degree-preserving randomized networks are obtained by shuffling connection among nodes while preserving nodes' degrees, they have no or very weak correlation between the degrees of connected nodes~\cite{PhysRevE.68.026112,PhysRevE.71.027103}. For such uncorrelated networks, the size of the viable core can be obtained analytically by solving self-consistent equations formulated in terms of the degree distributions as done in Refs.~\cite{PhysRevLett.96.040601,PhysRevLett.109.205703,LEE2023113645} or the adjacency matrix, called the message-passing method,  as done recently in Ref.~\cite{bianconi2023nature}. Here we take the former approach to understand the influence of the degree distribution on the core size in uncorrelated networks.

Following Ref.~\cite{PhysRevLett.109.205703},  let us first consider a compound node reached by following a link from a viable reaction node and denote the probability of the compound node to be inviable by $\alpha$. The compound node will be inviable if all the remaining neighboring reaction nodes,  except the viable one at the end of the followed link, are inviable. Next, let us consider a reaction node reached by following a link from a viable compound node, and denote the probability of the reaction node to be inviable by $\beta$.  The reaction node will be viable if (i) it is not removed by the random removal of $Q$ fraction of reaction nodes and (ii) all of its neighboring compound nodes are viable. These reasonings lead us to 
$\alpha = \sum_{k} \frac{k P_{\rm cmp}(k)}{\bar{k}} \beta^{k-1}$ and $1-  \beta = (1-Q)\sum_{q} \frac{q P_{\rm rxn}(q)}{\bar{q}} (1-\alpha)^{q-1}$, 
where $P_{\rm cmp}(k)\,[P_{\rm rxn}(q)]$ and $\bar{k}\, (\bar{q})$ are the degree distribution and the mean degree of compound (reaction) nodes, respectively. Introducing the generating functions $G_{\rm cmp}(z) \equiv \sum_{k=0}^\infty P_{\rm cmp}(k) z^k$ and 
$G_{\rm rxn}(z) \equiv \sum_{q=0}^\infty P_{\rm rxn}(q) z^q$, one can represent the previous equations as
\begin{align}
\alpha &= {G^\prime_{\rm cmp}(\beta)\over \bar{k}}, \nonumber\\ 
1-\beta &= (1-Q) {G^\prime_{\rm rxn} (1-\alpha) \over \bar{q}}.
\label{eq:gf}
\end{align}
Recalling that a reaction node is viable if it is not removed and all of its neighboring compound nodes are viable and a compound node is viable if it has at least two neighbor reaction nodes that are viable, one can find  the probability $v_{\rm rxn}\, (v_{\rm cmp})$ of a reaction (compound) node to be viable as
\begin{align}
    v_{\rm rxn} &= (1-Q)\sum_{q} P_{\rm rxn} (q) (1-\alpha)^q= (1-Q) G_{\rm rxn}(1-\alpha), \nonumber\\
    v_{\rm cmp} &=    \sum_{k\geq2} P_{\rm cmp} (k) \sum_{s=2}^{k} \binom{k}{s}(1-\beta)^{s}\beta^{k-s} \nonumber\\
  &= \sum_{k\geq2} P_{\rm cmp} (k) \left[1-\beta^k-k(1-\beta)\beta^{k-1}\right] \nonumber\\
  & = 1 - G_{\rm cmp}(\beta) - (1-\beta) G^\prime_{\rm cmp}(\beta).
   \label{eq:ccomprxn}
\end{align}
Finally, the relative  size $v$ of the viable core is given by 
\begin{equation}
v = n_{\rm cmp} v_{\rm cmp} + n_{\rm rxn} v_{\rm rxn} 
\label{eq:c}
\end{equation}
with $n_{\rm cmp}\, (n_{\rm rxn})$ being the fraction of compound (reaction) nodes, satisfying $n_{\rm cmp} + n_{\rm rxn}=1$,  in the considered metabolic network.

The solutions to Eqs.~\eqref{eq:gf}, \eqref{eq:ccomprxn}, and \eqref{eq:c} are dependent only on the degree distributions $P_{\rm cmp}(k)$ and $P_{\rm rxn}(q)$. The degree distributions averaged over species 
are shown in Figs.~\ref{fig:robust}(c) and~\ref{fig:robust}(d).  
The broad degree distributions for compounds are universal, which leads the viable cores to shrink gradually as $Q$ increases as shown in Fig.~\ref{fig:robust}(a)~\cite{PhysRevLett.109.205703,PhysRevLett.96.040601}. Such a gradual shrink may  not be the case when the degree distribution is narrow and the mean degrees are sufficiently large. In the completely randomized networks that are obtained by assigning links to randomly selected node pairs and thus preserve the total number of links but do not preserve the degree sequences, the viable core size drops abruptly to a small value at a critical value $Q_c$ [Fig.~\ref{fig:robust}(a)], denoted by `link-preserving'. With the degree distributions given in the Poissonian form, one can find that such discontinuous transitions can occur if the mean degrees are larger than $e=2.718182\ldots$, i.e., $\bar{k}>e$ and $\bar{q}>e$~\cite{PhysRevLett.109.205703,PhysRevLett.96.040601}. See Appendix~\ref{seca:coresize} for derivation. We find 5463/5469 $\simeq$ 99.8\% species satisfy this condition.

\begin{figure}
\includegraphics[width=\columnwidth]{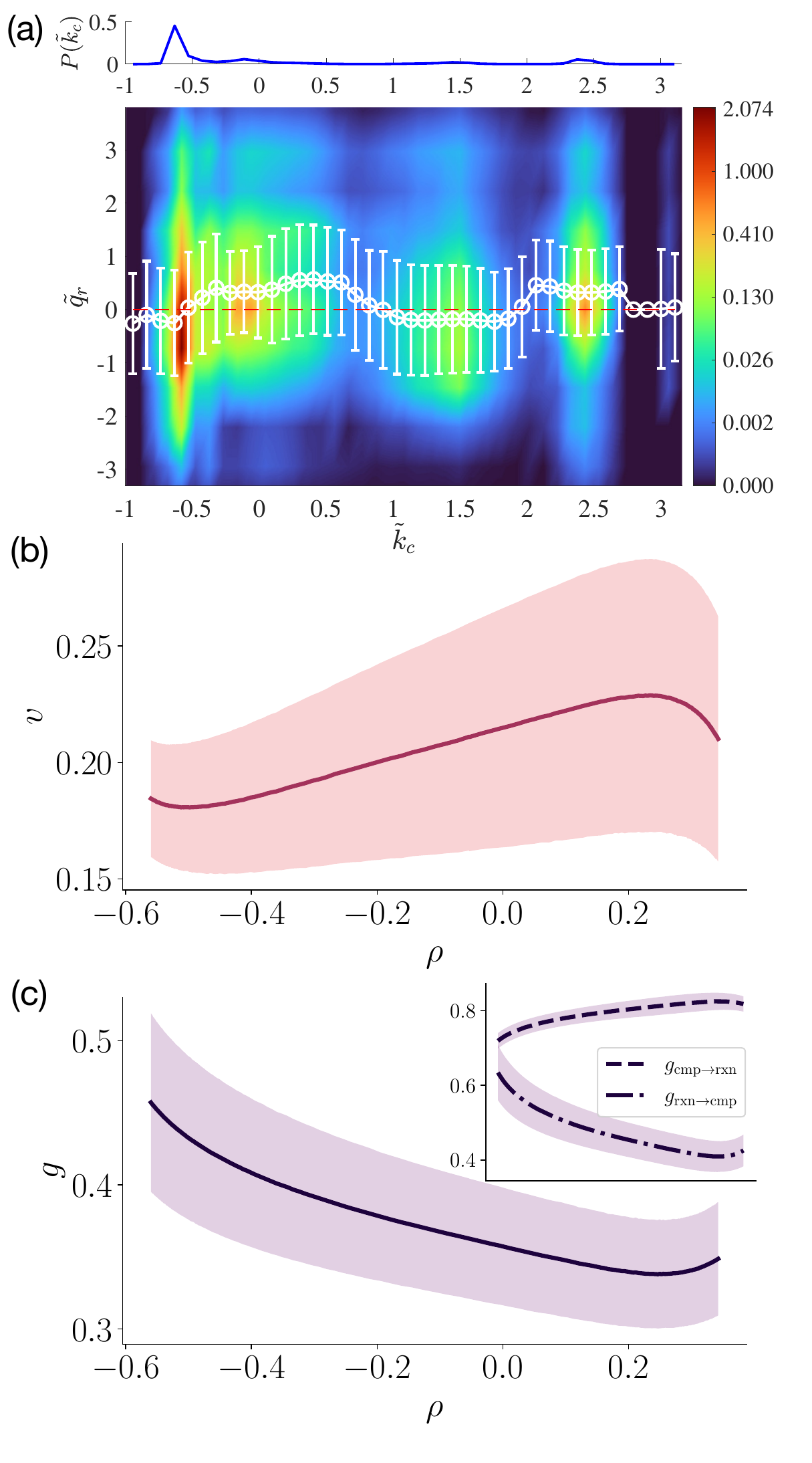}
\caption{Effects of degree-degree correlations on the size of viable cores. 
(a) Positive correlations in empirical networks. The color plot of the joint pdf $P(\tilde{k}_c,\tilde{q}_r)$ of the standardized degrees of the compound node $c$ and the reaction node $r$ at both ends of a link, averaged over species, is shown. The conditional mean $\langle \tilde{q}_r\rangle(\tilde{k}_c)=\int d\tilde{q}_r \tilde{q}_r P(\tilde{q}_r|\tilde{k}_c)$ is represented by white circles along with the standard deviation represented by white bars. In the range $-1\lesssim \tilde{k}_c\lesssim 0$ where the pdf $P(\tilde{k}_c)$, shown in the upper part, is large, the increase of $\langle \tilde{q}_r\rangle(\tilde{k}_c)$ with $\tilde{k}_c$ is noticeable.
(b) The relative size $v$ of the viable core in the degree-preserving correlated networks of given assortativity $\rho$. The solid line represents the mean and the shade represents the standard deviation.  
(c) Branching ratio $g$ in the correlated networks as a function of assortativity. The solid lines represent the mean values and the shades the standard deviation. Inset: The compound-to-reaction branching ratio $\gcr$ and the reaction-to-compound branching ratio $\grc$ versus $\rho$. 
}
\label{fig:correlation}
\end{figure}

\section{Origin of enhanced viable cores}
\label{sec:correlation}

The significantly larger viable cores than expected in the degree-preserving randomized networks suggest that correlations or higher-order structural characteristics beyond the degree sequences play a crucial role in the formation of the viable core. In this section, we explore the correlation of the degrees of connected nodes in real metabolic networks and study how they can affect the viable core and the possible origin of such correlations. 

\subsection{Positive degree-degree correlations}

Our analysis shows that the degrees $q$ and $k$ of the end nodes, reaction- and compound-type respectively, of a link are positively correlated in the empirical networks. Their Pearson correlation coefficient or {\it assortativity} is defined for each species $s$ as ~\cite{PhysRevLett.89.208701},
\begin{equation}
\rho^s = \sum_{c,r} {A_{cr}^s \over L^s} \left({k_c^s - k_{\rm nn}^s\over \sigma_{\rm cmp}^s}\right) \left({q_r^s - q_{\rm nn}^s \over \sigma_{\rm rxn}^s}\right),
\label{eq:assortativity}
\end{equation}
where $A_{cr}^s$ is the adjacency matrix of the metabolic network, and $k_{\rm nn}^s \equiv \sum_{c,r} (A_{cr}^s/L^s) k_c^s$ and $q_{\rm nn}^s \equiv \sum_{c,r} (A_{cr}^s/L^s) q_r^s$ are the mean degrees of the end nodes of a link, and $\sigma_{\rm cmp}^s \equiv [\sum_{c,r} (A_{cr}^s/L^s) (k_c^s - k_{\rm nn}^s)^2]^{1/2}$ and $\sigma_{\rm rxn}^s \equiv [\sum_{c,r} (A_{cr}^s/L^s) (q_r^s - q_{\rm nn}^s)^2]^{1/2}$ are the standard deviations. The assortativity of the real metabolic networks is given by $\rho_{\rm emp}\simeq 0.13 \pm 0.02$ and significant with P values less than 0.01 for almost all species.  
Also we compute the probability density function (pdf) of the standardized degrees $\tilde{k}_c^s = {k_c^s - k_{\rm nn}^s\over \sigma_{\rm cmp}^s}$ and $\tilde{q}_r^s = {q_r^s - q_{\rm nn}^s \over \sigma_{\rm rxn}^s}$, and the conditional mean $\langle \tilde{q}^s\rangle (\tilde{k}) = \sum_{c,r} (A_{cr}^s/L^s) \tilde{q}_r^s \delta(\tilde{k}_c^s-\tilde{k})/\sum_{c,r} (A_{cr}^s/L^s)\delta(\tilde{k}_c^s-\tilde{k})$. Note that the assortativity is one of the moments of the pdf,  $\rho =\int d\tilde{k} d\tilde{q} P(\tilde{k}, \tilde{q}) \tilde{k} \tilde{q}$.  In Fig.~\ref{fig:correlation}(a) are shown the pdf and the conditional mean averaged over species. One can see that $\langle \tilde{q}\rangle (\tilde{k})$ grows with $\tilde{k}$ in the most probable range $-1\lesssim\tilde{k}\lesssim 0$ supporting the positive correlation between $\tilde{q}$ and $\tilde{k}$.

To scrutinize the influence of the degree-degree correlation on the size of the viable core, we generate an ensemble of artificial degree-preserving {\it correlated} networks that exhibit a given assortativity. To be specific, for given species $s$, we perform the Monte-Carlo sampling of different configurations---different adjacency matrices $A_{cr}$---while preserving the given degree sequence by swapping the end nodes of two links with a  Hamiltonian $\mathcal{H}=-J\sum_{c, r} A_{cr} k_{c}^s q_{r}^s$. The assortativity is measured in the equilibrated networks for each $J$.
Note that the degree-preserving randomized networks are generated with $J=0$.  The relative size $v$ of the viable cores of these degree-preserving correlated networks, averaged over species, increases as assortativity $\rho$ increases over a wide range of $\rho$, i.e., $-0.5\lesssim \rho\lesssim 0.2$, which includes most of the empirical values of assortativity [Fig.~\ref{fig:correlation}(b)]. This demonstrates the contribution of the positive degree-degree correlations to the enhancement of the viable cores in the empirical networks. Yet, we also note that they are not the only cause of the core enhancement;  The size of the empirical viable cores, $v = 0.4\pm 0.04$, is larger than $v = 0.22\pm 0.02$ of those artificial correlated networks at the empirical values of assortativity $\rho_{\rm emp}$.

To understand why a positive (negative) correlation makes a larger (smaller) viable core for a given degree sequence, we investigate the branching ratio $g$ of inviable compounds between adjacent steps of pruning; If there are $C_1$ inviable compounds in the current step and after removing them and their connected reactions nodes, there appear $C_2$ newly inviable compounds in the next step, the branching ratio will be given by $g\equiv{C_2 \over C_1}$.  One can expect a larger (smaller) core with a smaller (larger) branching ratio. To see the dependence of the branching ratio on assortativity, we decompose the branching ratio into the compound-to-reaction one and the reaction-to-compound one as 
\begin{equation}
g = \gcr  \grc \\
\ {\rm with} \ \gcr \equiv {R_1\over C_1} \ {\rm and } \ \grc \equiv {C_2 \over R_1},
\label{eq:gdecomp}
\end{equation}
where $R_1$ is the number of the current-step inviable reaction nodes, those connected to the $C_1$ current-step inviable nodes,  and $C_2$ is the number of the next-step inviable nodes that will be made inviable only in the next step by pruning in the current step. 

If a network has a negative degree-degree correlation, the compound-to-reaction branching ratio $\gcr$ can be small; Inviable compounds may be concentrated, being connected to a few hub reaction nodes. Each of those current-step inviable reaction nodes has many neighbor compounds and thus can leave many next-step inviable compounds, resulting in large $\grc$. In contrast, $\gcr$ cannot be so small in a network with a positive degree-degree correlation since inviable compounds are distributed, each being connected to each different reaction node of a small degree. Those current-step inviable reaction nodes have small degrees and thus each will leave a small number of next-step inviable compounds, resulting in small $\grc$. 

These reasonings can be summarized as follows: As assortativity increases, $\gcr$ increases but $\grc$ decreases. This is confirmed in our artificial correlated networks. In the inset of Fig.~\ref{fig:correlation}(c), we show  the two branching ratios obtained by computing $C_1, R_1$, and $C_2$ in the early-stage of pruning in each correlated network and averaging over species, which behave as functions of $\rho$ as expected within a range of interest. A crucial observation is that the increase of $\gcr$ with $\rho$ is much weaker than the decrease of $\grc$ with $\rho$.  Such quantitative difference in the dependence on $\rho$ results in the decrease of $g$, the product of the two branching ratios, with increasing $\rho$; $g \simeq 0.4$ at $\rho = -0.4$ while $g\simeq 0.35$ at $\rho=0.1$.  

Such dependencies of the branching ratios and the core size on assortativity help us understand how a structural correlation can influence the viable core. We should remark that the behaviors of $v(\rho)$ and $g(\rho)$ may be changed in extremely correlated networks; $v$ decreases as $\rho$ increases in the range $\rho\lesssim -0.5$ and $\rho\gtrsim 0.2$ [Fig.~\ref{fig:correlation}(b)] and $g$ increases with $\rho$ in the range $\rho\gtrsim 0.2$ [Fig.~\ref{fig:correlation}(c)]. The investigation of these deviations may be interesting and need further investigation. We also note that the empirical branching ratio $g=0.2\pm0.03$ is smaller than the ratio $g\simeq 0.34$ from the artificial correlated networks at the empirical value of assortativity, implying again the influence of higher-order structure on the viable core and the branching ratio of pruning. 

\begin{figure}
\includegraphics[width=\columnwidth]{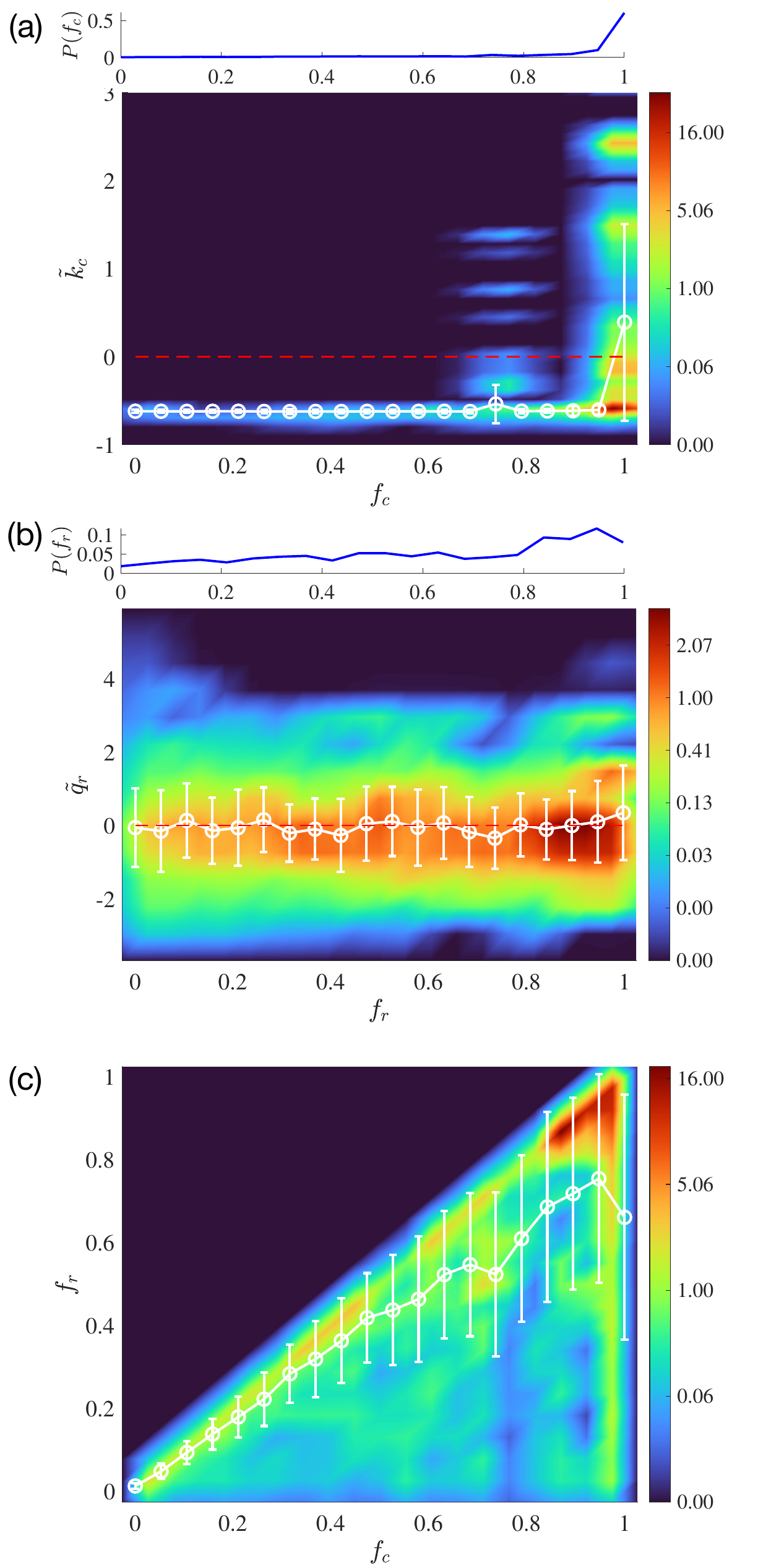}
\caption{Correlation of the degree and frequency of compounds and reactions in real metabolic networks. 
(a) The color plot of the joint pdf $P(f_c,\tilde{k}_c)$ of the frequency and the standardized degree of the compound node $c$ at one end of a link, averaged over species, is shown. The conditional mean $\langle \tilde{k}_c\rangle(f_c)=\int d\tilde{k}_c \tilde{k}_c P(\tilde{k}_c|f_c)$ is represented by white circles along with the standard deviation represented by white bars. $\langle \tilde{k}_c\rangle(f_c)$ increases abruptly with $f_c$ in the range $0.9\lesssim f_c\lesssim 1$ in which the pdf $P(f_c)$ is large.
(b) The joint pdf $P(f_r,\tilde{q}_r)$ of the frequency and the standardized degree of the reaction node $r$ at one end of a link is shown, along with the conditional mean $\langle \tilde{q}_r\rangle(f_r)$. In the range $\tilde{q}_r\gtrsim 0.8$ with large $P(\tilde{q}_r)$, $\langle \tilde{q}_r\rangle(f_r)$ exhibits a noticeable increase with $f_r$. 
(c) The joint pdf $P(f_c,f_r)$ of the compound and the reaction at both ends of a link. The conditional mean $\langle f_r\rangle(f_c)$ increases persistently with $f_c$.}
\label{fig:pop}
\end{figure}

\subsection{Evolutionary origin of positive correlations}

Given that a positive degree-degree correlation can enlarge the viable core and thereby enhance stability, one can suspect that an evolutionary pressure might have been imposed towards generating the positive correlations in the real metabolic networks during evolution.  While it is a simple and compelling scenario, here we present evidence that a neutral evolution of metabolism, without an evolutionary pressure, over a growing tree of species on a long-time scale can generate the positive correlations.

The frequency $f_r$ ($f_c$) of a reaction $r$ (compound $c$) to be present in a species' metabolism among species has been found to be so heterogeneous ~\cite{Bernhardsson:2011ai,Kim:2015aa,Kim:2019aa}  that its distribution takes a power-law form with exponent one~\cite{Kim:2015aa,Kim:2019aa,PhysRevLett.132.018401}. It has been shown in Ref.~\cite{PhysRevLett.132.018401} that such heterogeneous frequencies of reactions and compounds may originate from their different times of the first appearance in the metabolism of a living species and their inheritance to the descendant species during the co-evolution of metabolism and species tree.  In this framework, the early-recruited compounds and reactions have high frequency and moreover, are expected to have larger degrees in the metabolic networks of the species containing them since they have had more chances to be connected to other nodes during evolution than the late-recruited ones. Therefore we can think of the empirical positive degree-degree correlations as originating from the co-recruitment of connected reactions and compounds in the evolving metabolism in a growing species tree.  

The empirical data support this expectation. The Pearson correlation coefficient between the frequency $f_c$ and the degree $k_c^s$ of a compound at the end of a link in a species is $0.34
\pm 0.03$  (P $\lesssim 4\times 10^{-6}$ for all species)  with the fluctuation arising among species, demonstrating their significant correlations. In each species, the end compound node of a link is likely to have high frequency and such high-frequency compounds show clearly a positive correlation between the standardized degree $\tilde{k}_c$ and the frequency $f_c$ as shown by their joint pdf and the conditional mean $\langle \tilde{k}_c\rangle (f_c)$ [Fig.~\ref{fig:pop}(a)]. The correlation between $f_r$ and $q_r^s$ of a reaction is $0.06 \pm 0.06$, which is weaker than the correlation for compounds' degree and frequency but still significant for many species (P $\lesssim 0.05$ for $4130/5469\approx 76\%$ species). The pdf $P(f_r, \tilde{q}_r)$ and the conditional mean $\langle q_r\rangle(f_r)$ show that the correlation is particularly significant for reactions of high frequency in Fig.~\ref{fig:pop}(b). 

Also, as expected, the frequencies of a connected pair of reaction and compound nodes are strongly correlated; Their Pearson correlation coefficient is  $0.42 \pm 0.04$ (P value $\lesssim 0.0015$ for all species) and the pdf $P(f_c, f_r)$, and the conditional mean $\langle f_r\rangle(f_c)$ clearly show the correlation in Fig.~\ref{fig:pop}(c).  Collecting all these results,  we find that for a connected pair of a compound node and a reaction node, their degrees are positively correlated with their frequencies respectively and their frequencies are very similar, as expected from the metabolism evolution model as sketched above. All these factors result in the positive correlations of the degrees of the pair and, in turn, contribute to the enhancement of the viable core.

\section{Summary and Discussion}
\label{sec:summary}

Metabolism serves to perform key life processes robustly and at the same time interacts closely with fluctuating environments, and it is an important question how finite cellular resources are divided between such opposite tasks.  By the graph-theoretic approach, we have investigated the viable core of the metabolic network, expected to function stably and contribute to homeostasis, across thousands of species to discover that those cores are larger than would be observed if metabolic compounds and reactions were paired randomly. We have presented evidence that the positive degree-degree correlation of a connected pair of a reaction and a compound, identified empirically, can be a cause of those enhanced viable cores.  We have also shown that the co-recruitment of reactions and compounds into the metabolism over a growing phylogenetic tree of species during evolution can give rise to such positive correlations.

Here we have focused on aggregate properties, and therefore extending our study to assessing the structural stability of individual elements of metabolism, reactions and compounds, and individual species' metabolism. We have identified the strikingly different behaviors in the decay of the viable core under node removal between homogeneous and heterogeneous networks and its analytic understanding can be desirable. It will be also interesting to extend our study to directed bipartite metabolic networks obtained by considering the irreversibility of metabolic reactions.

\begin{acknowledgments}
This work was supported by grants from the National Research Foundation of Korea (NRF) funded by the Korean Government [No. NRF-2021R1C1C1007918 (M.J.L.) and NRF-2019R1A2C1003486 (D.-S.L.)], and a KIAS Individual Grants [No. CG079902(D.-S.L) and No. CG074102 (S.Y.)] from Korea Institute for Advanced Study. We are grateful to the Center for Advanced
Computation in KIAS for providing computing resources. 
\end{acknowledgments}


%

\appendix
\section{Size of  the viable core in completely random bipartite networks under random failure: Discontinuous or continuous transition}
\label{seca:coresize}

Here we study the viable core of the random bipartite networks consisting of the compound- and reaction-type nodes, which have the Poisson degree distributions $D_{\rm cmp}(k) = {\bar{k}^k \over k!} e^{-\bar{k}}$ and  $D_{\rm rxn}(q) = {\bar{q}^q \over q!} e^{-\bar{q}}$ with $\bar{k}$ and $\bar{q}$ the mean degrees of both types of nodes. Using their generating functions $G_{\rm cmp}(z) = e^{-(1-z)\bar{k}}$ and $G_{\rm rxn}(z) = e^{-(1-z)\bar{q}}$ in Eq.~\eqref{eq:gf}, we obtain 
\begin{equation}
\alpha = e^{-\bar{k} (1-\beta)} \ {\rm and} \ 1-\beta = (1-Q) e^{-\bar{q} \alpha}.
\label{eqa:gf}
\end{equation}
The solution $\alpha$ and $\beta$ to Eq.~\eqref{eqa:gf} are then used in Eqs.~\eqref{eq:ccomprxn} and \eqref{eq:c} to give the relative size of the viable core. 

One can establish from Eq.~\eqref{eqa:gf} 
\begin{equation}
\alpha = f(\alpha) \ {\rm with} \ f(\alpha)\equiv  e^{-\bar{k} (1-Q) e^{-\bar{q} \alpha}}.
\label{eqa:f}
\end{equation}
The intersection of $y=f(\alpha)$ and $y=\alpha$ in the $(\alpha,y)$ plane gives the solution $\alpha$.
Note that a small (large) value of $\alpha$ leads to a large (small) core.   
The function $f(\alpha)$ increases monotonically with $\alpha$. For sufficiently large $\bar{k}$ and $\bar{q}$, one can see that $y=f(\alpha)$ is convex ($f^{\prime\prime}(\alpha)>0$) for small $\alpha$ and concave ($f^{\prime\prime}(\alpha)<0$) for large $\alpha$, and meets $y=\alpha$ at one, two, or three points depending on $Q$ as follows.  i) When $Q$ is small, $y=f(\alpha)$ intersects $y=\alpha$ at three points with the leftmost intersection point, giving the solution $\alpha$. ii) At $Q=Q_*$, $y=f(\alpha)$ is tangential to $y=\alpha$ at $\alpha_*$. iii) For $Q>Q_*$, they intersect once, close to $\alpha=1$. Consequently,  the solution $\alpha$ increases from a small value with increasing $Q$ and jumps to a large value at $Q=Q_*$, leading to the discontinuous drop in the core size at a threshold $Q_*$ as seen in Fig.~\ref{fig:robust}(a). When the mean degrees are not large enough, $y=f(\alpha)$ meets $y=\alpha$ only once for all values of $Q$, and one can see a gradual increase of $\alpha$ and a gradual decrease of the core size $c$ with increasing $Q$. The threshold for such discontinuous transitions can be identified by noting that at the threshold, the curve $y=f(\alpha)$ is tangential to $y=\alpha$ at $\alpha_*$ and its convexity also changes at $\alpha_*$. Therefore, at the threshold, we find three relations $\alpha_* = f(\alpha_*), 1 = f^\prime(\alpha_*)$, and $0 = f^{\prime\prime}(\alpha_*)$ are satisfied. Using Eq.~\eqref{eqa:f}, we find that the relations are satisfied when $\alpha_* = e^{-1}$,  $\bar{q} = e$, and  $\bar{k}={e\over 1-Q_*}$.  The last expression implies that a solution to $Q_*$ can exist between $0$ and $1$ if $\bar{k}>e$. Therefore we can see that a discontinuous transition in the core size as a function of $Q$ occurs when the mean degrees are larger than $e$, i.e., 
\begin{equation}
\bar{k}>e \,  {\rm and} \,  \bar{q}>e.
\end{equation}

\section{Generating degree-preserving correlated networks}
\label{seca:mc}


To explore the role of degree-degree correlations in forming a viable core, we generate networks by 
changing the neighboring of nodes in the original networks while preserving the degree of each node. 
The preservation of degree sequences makes the moments of degrees constant, resulting in the constant moment-relevant quantities $k_{\mathrm{nn}}^s$, $q_{\mathrm{nn}}^s$, $\sigma_{\mathrm{cmp}}^s$, and $\sigma_{\mathrm{rxn}}^s$ in Eq.~\eqref{eq:assortativity} in all the generated networks for a given species $s$, while the term $\sum_{c,r}A^s_{cr}k^s_c q^s_r$ may be different from network to network.

Therefore, to control the degree-degree correlation in the generated networks, we adopt the Hamiltonian suggested in Ref.~\cite{PhysRevE.76.026116}, which for the bipartite metabolic network is formulated as
\begin{equation}
\mathcal{H}=-J\sum_{c, r} A_{cr} k^s_{c} q^s_{r},
\end{equation}
where $J$ is the coupling strength, $k^s_c$ and $q^s_r$ are the degrees of a compound $c$ and a reaction $r$ for a given species $s$, respectively, and $A_{cr}=1$ if a chemical reaction $r$ is associated with a compound $c$ (schematically $c \to r$), or $A_{cr}=0$ otherwise ($c \not\to r$).

We generated an ensemble of artificial correlated networks for a given $J$ by conducting a degree-preserving Monte Carlo (MC) simulation, which involves the swapping of randomly chosen pairs of links. For example, consider a scenario where a pair of links $a$ and $b$, namely,  $c_a \to r_a$ and $c_b \to r_b$ are randomly selected, where $c_a \not\to r_b$ and $c_b \not\to r_a$. The swapping can be executed as $c_a \to r_b$ and $c_b \to r_a$ with $c_a \not\to r_a$ and $c_b \not\to r_b$. This swap is accepted with the probability  $\min \left\{1, e^{-(\mathcal{H}^\prime-\mathcal{H})}\right\}$, where $\mathcal{H}^\prime$ is the Hamiltonian after the swap is implemented.

Technically, we sample an ensemble of a species for a given $J$ as 10 networks every 5 MC steps after stationarity is reached, where link swapping occurs for the total number of links in a single MC step. Consequently, each data point in Figs.~\ref{fig:correlation}(c) and (d) is averaged over 10 configurations per species and across all species.

\end{document}